\documentclass[aps,prl,twocolumn,groupedaddress,showpacs,floatfix]{revtex4}

\usepackage[latin1]{inputenc}
\usepackage[english]{babel}
\usepackage{setspace, graphicx, amssymb}
\usepackage{amsmath}
\usepackage{amscd}
\usepackage{rotating}
\usepackage{color}
\usepackage{epstopdf}
\usepackage{subfigure}

\newcommand{\tr}{{\rm Tr\thinspace}}

\newcommand{\abs}[1]{\left\vert #1 \right\vert}
\newcommand{\expect}[1]{\left\langle{#1}\right\rangle}

\newcommand{\dd}[1]{\frac{d}{d#1}}
\newcommand{\timeorder} {\underset{\leftarrow}{\mathcal{T}}}

\newcommand{\diag}[1]{\mathrm{diag}\{#1\}}

\newcommand{\bopt}{\vec b_{\mathrm{opt}}}
\newcommand{\lambdaopt}{\vec \lambda_{\mathrm{opt}}}

\begin{document}

\title{Fighting dephasing noise with robust optimal control}

\author{Kevin C. Young\textsuperscript{1,3}$^\dagger$ }
\email{kcyoung@berkeley.edu}
\author{Dylan J Gorman\textsuperscript{1}$^\dagger$ }
\author{K. Birgitta Whaley\textsuperscript{2,3} }
\affiliation{\vspace{.1cm}
	\textnormal{{$^\dagger$These authors contributed equally to this work.} }
	\vspace{.1cm}\\
	\mbox{$^1$Department of Physics, University of California, Berkeley, California 94720, USA}\\
	\mbox{$^2$Department of Chemistry, University of California, Berkeley, California 94720, USA}\\
	\mbox{$^3$Berkeley Quantum Information and Computation Center}\\
	\vspace{5pt}}
\date{\today}

\begin{abstract}
We address the experimentally relevant problem of robust mitigation of dephasing noise acting on a qubit.  We first present an extension of a method for representing $1/\omega^{\alpha}$ noise developed by Kuopanportti et al.~\cite{kuop2008} to the efficient representation of arbitrary Markovian noise.  We then add qubit control pulses to enable the design of numerically optimized, two-dimensional control functions with bounded amplitude, that are capable of decoupling the qubit from the dephasing effects of a broad variety of Markovian noise spectral densities during arbitrary one qubit quantum operations.  We illustrate the method with development of numerically optimized control pulse sequences that minimize decoherence due to a combination of $1/\omega$ and constant offset noise sources.  Comparison with the performance of standard dynamical decoupling protocols shows that the numerically optimized pulse sequences are considerably more robust with respect to the noise offset, rendering them attractive for application to situations where homogeneous dephasing noise sources are accompanied by some extent of heterogeneous dephasing.  Application to the mitigation of dephasing noise on spin qubits in silicon indicates that high fidelity single qubit gates are possible with current pulse generation technology.
\end{abstract}

\maketitle

\section{Introduction}
Coherent control of interactions between quantum bits (qubits) and their environment is an essential component of the search for realization of reliable quantum computation.  Exercising control of coherence by application of carefully designed pulse sequences is a standard tool in NMR and ESR, where particularly powerful sets of tools have been developed for protection against qubit dephasing. Such coherent control provides a complementary approach to the protection of quantum information by encoding, whether with active quantum error correction codes or passive encoding into decoherence free ('noiseless') subspaces and subsystems.   In the quantum information processing community, the application of coherent control ideas to preservation of qubit coherence has prompted an explosion of work in the field of dynamical decoupling \cite{khod2007, uhrig2007} which builds on the classic spin echo techniques from NMR.  An alternative approach is the design of numerically optimized control pulse sequences, which have the advantage of flexibility and ready applicability to both quantum memory and protection of arbitrary quantum gates against  dephasing and leakage errors ~\cite{kuop2008, mott2006, wilhelm2009}.   

In this work we consider the design of optimal pulse sequences for control of decoherence during single qubit operations when the qubit is coupled to source of Markovian noise that possesses an arbitrary noise spectrum.
The Hamiltonian governing the evolution of the qubit is taken to be
	\begin{equation}
	\label{hamiltonian}
		 H(t) = \frac12\left(a_x(t)\sigma_x+ a_y(t)\sigma_y +\eta(t) \sigma_z \right),
	\end{equation}
where we are working in a rotating frame so that the qubit energy level splitting is, on average, zero. 
Here, $a_x(t)$ and $a_y(t)$ are bounded-amplitude control fields, while $\eta(t)$ is a classical stochastic process.   In the absence of control, this Hamiltonian generates dephasing of the qubit, at a rate that depends on both the amplitude distribution and the temporal correlation function of the stochastic process $\eta(t)$.  
Previous work in our group has modeled this stochastic process as a multi-state Markovian fluctuator, the dynamics of which may be tuned to mimic a particular noise source~\cite{kuop2008}
The use of a multi-state Markovian fluctuator allows the evolution under the Hamiltonian, Eq.~\eqref{hamiltonian} to be efficiently solved through use of a deterministic master equation \cite{saira2007}.  
In the present work, we extend this approach from the $1/\omega^{\alpha}$ spectral noise sources for which analytic representations of the fluctuator dynamics could be found~\cite{kuop2008}, to representation of a broad variety of Markovian spectral noise distributions by making use of convex optimization techniques.  We then use gradient ascent methods as in Ref.~\cite{kuop2008} to derive control sequences for single qubit operations.  Following established literature convention, these pulse sequences will be generically refered to as ``GRAPE sequences'' (i.e., gradient-ascent pulse engineering sequences) \cite{khaneja2005}.  We focus here on two specific single qubit operations: i) the identity, which is equivalent to extension of qubit coherence, i.e., to quantum memory; ii) the Hadamard gate.  
We analyze the robustness of these pulse sequences for protection against the combined effects of $1/\omega$ noise and zero frequency noise (the latter is equivalent to a resonance frequency error) and compare with the corresponding performance of standard dynamical decoupling sequences, in particular with Carr--Purcell sequences.   We find that the numerically optimized control sequences improve on the dynamical decoupling sequences over a broad range of zero frequency noise offsets, resulting in considerably greater robustness in addition to improved decoherence mitigation.  Finally, we demonstrate the applicability of the method for current day experiments by making explicit application to the protection of coherence for dopant spin qubits in silicon using realistic estimates of spectral noise and control pulse capabilities.  The results indicate that gates with errors less than $10^{-5}$ can be designed and implemented with current technology.  This is well below current estimates of the  fault-tolerant threshold \cite{stean2003}.

\section{Simulated Noise Model}

Qubit evolution under classical noise $\eta(t)$ is simulated using an extension of a method first developed in \cite{kuop2008}, which we review and then expand upon here.  We consider $\eta(t)$ in Eq.~\eqref{hamiltonian} to represent a multi-state Markovian fluctuator having $N$ noise states.  The noise state $k$ has amplitude $\eta_k$ and occupation probability $p_k(t)$ at time, $t$.  These noise amplitudes and occupation probabilities will be represented as vectors, $\vec {\eta}$ and $\vec p$.  Transitions between noise states are governed by a rate matrix, $\Gamma$,
	\[ \frac{d}{dt}\vec p(t) = \Gamma \, \vec p(t). \]
To conserve probability, the transition rate matrix must satisfy $\sum_k \Gamma_{kj} = 0$.  This constraint implies that the vector $\vec p_{\rm{s}} = (1/N,1/N,\ldots,1/N)$ is a stationary probability vector, and is an eigenvector of $\Gamma$ with zero eigenvalue.  We shall limit our study of the rate matrices to those satisfying the additional requirement,  $\Gamma = \Gamma^{\rm{T}}$. This condition makes the forward and backward transition rates between any two noise states to be equal, enforcing time-reversal invariance on the fluctuator dynamics.  

The noise source may be further characterized by its power-spectral density, $S(\omega)$. 
\begin{equation}
	\label{powerspec}
	S(\omega) = \int  C(t) e^{-i\omega t}dt.
\end{equation}
The power spectrum of the multistate fluctuator is calculated through its temporal correlation function
	\begin{align*}
		 C(t) 	&= \expect{\eta(t)\eta(0)} \\
		 	&= \sum_{i,j} P(\eta_j, t \vert \eta_i, 0) \eta_j P(\eta_i(0)) \eta_i \\
			&= \frac{1}{N} \sum_{i,j} \eta_i \left[e^{\Gamma \abs{t}} \right]_{ij} \eta_j,
	\end{align*}
where $P(\eta_i(0))\equiv p_i(0)$ denotes the probability of the fluctuator being in state $i$ at time $t=0$ and $P(\eta_j, t \vert \eta_i, 0)$ the conditional probability of it being in state $j$ at time $t$, given state $i$ at $t=0$.  Here we have chosen as the initial noise probabilities the stationary vector $\vec{P}(t) = \vec{p}_s = ( 1/N,1/N,\ldots,1/N)$.  Because $\Gamma$ is a symmetric matrix, it can be diagonalized by an orthogonal matrix, $\Gamma = V^\dagger \Lambda V$, so that
	\begin{equation}
	\label{correlation}
		 C(t) = \frac{1}{N} \vec\eta^{\,\dagger} V^\dagger e^{\Lambda \abs{t}} V \vec \eta = \vec b^{\, \dagger} e^{\Lambda \abs{t}} \vec b,
	\end{equation}
where we have defined the transformed noise amplitude vector $\vec b = V \vec \eta /\sqrt{N}$ and $\Lambda = \diag{\lambda_1, \lambda_2, \ldots, \lambda_N}$ is the diagonal matrix of eigenvalues of $\Gamma$.  For convenience, the eigenvalues are ordered $i > j \Rightarrow \lambda_i<\lambda_j$.  The corresponding power-spectral density, Eq.~\eqref{powerspec}, is a sum of zero-mean Lorentzian distributions:
	\begin{equation}
		\label{lorentzian}
		S\!\left(\omega ; \vec{\lambda}, \vec b \right) = \sum_j \frac{-2\,  b_j^2 \lambda_j}{\lambda_j^2 + \omega^2}
	\end{equation}
Ref.~\cite{kuop2008} derived an analytic form of $\Gamma$ and $V$ that generates noise with a $1/\omega^{\alpha}$, $0< \alpha <2$, power spectrum.  As noted there, numerical optimization may result in a more accurate representation.  

As with all Markovian processes, the form of Eq.~\eqref{lorentzian} is, in accordance with Doob's theorem \cite{doob1942}, a sum of Lorentzians.  This form constrains the possible target spectra to those which are monotonically decreasing and which never decay faster than $1/\omega^2$.  We have found that by proper choice of $\vec\lambda$ and $\vec b$, this spectrum may indeed be brought arbitrarily close to a given target spectrum, $S_t(\omega)$, (chosen with the above constrains in mind) over a finite specified range of frequencies, $\omega \in [\omega_{\rm{min}}, \omega_{\rm{max}}]$.   The choice of $\vec \lambda$ and $\vec b$ is made by a numerical optimization that minimizes the deviation of Eq.~\eqref{lorentzian} from the target spectrum. In particular, we carry out the following optimization:
	\begin{align*}
		\underset{\vec\lambda, \vec b}{\mathrm{minimize}} &\hspace{.5cm} \int_0^\infty  W(\omega)
				\left(S\!\left(\omega ; \vec{\lambda}, \vec b\right) -  S_t(\omega) \right)^2 d\omega \\
		\mathrm{subject \,to} & \hspace{.5cm} b_i \ge 0,\hspace{.2cm}  \lambda_i \le 0.
	\end{align*}
Since i) $\omega$ can span many orders of magnitude, and ii) analytic representations of power spectra often  diverge at $\omega = 0$, we have incorporated here a weighting function, $W(\omega)$, into the usual $L^2$ distance measure.  In particular, we have set $W(\omega) = 1/\omega$ for $\omega\in [\omega_{\min}, \omega_{\rm{max}}]$ and $W(\omega) = 0$ otherwise.  This weight function is uniformly distributed in $\log\omega$,
 preventing higher frequencies from dominating the integral.   Restricting ourselves to the range $\lambda_i \le 0$ is physically realistic, since positive eigenvalues would not conserve probability and would cause the correlation function Eq.~\eqref{correlation} to diverge at long times.    
 
The results of this optimization are the two vectors $\bopt$ and $\lambdaopt$.  Recall that the constraints on $\Gamma$ imply the existence of a stationary probability vector, $\vec p_{\rm{s}}$ with eigenvalue $\lambda_1=0$.  This implies that one component of $\bopt$, say $b_1$, can be taken to be a free parameter and may be chosen to make the arithmetic mean of $\vec\eta=\sqrt{N}V^\dagger \vec b$ equal to zero, guaranteeing the existence of the stationary solution $\vec{p}_s = ( 1/N,1/N,\ldots,1/N)$.
We note that it is convenient to further make a restriction to $b_i\ge 0$ during the numerical optimization, because the power spectrum depends only on $b_i^2$.  However, following the optimization, we may subsequently adjust the signs of all components $b_{i\ne 1}$ so that $\abs{b_1}$ is as small as possible, consistent with the existence of the stationary probability vector.

It now remains to construct a valid transition rate matrix $\Gamma$ with eigenvalues given by $\lambdaopt$.  This is again done by a numerical optimization, namely
	\begin{align*}
		\underset{ \Gamma  }{\mathrm{minimize}} &\hspace{.5cm}   \left(\mathrm{eigs}(\Gamma) - \vec\lambda_{\rm opt} \right)^2 \\
		\mathrm{subject \,to} & \hspace{.5cm}  \Gamma = \Gamma^\mathsf{T}\!,\hspace{.2cm}  \Gamma_{i\ne j} \ge0\\
						& \hspace{.5cm} \sum_{j} \Gamma_{ij} =0,
	\end{align*}
with $\mathrm{eigs}(\Gamma)$ the vector of eigenvalues of $\Gamma$.  

Both of the above optimizations can performed very quickly using standard 
convex optimization tools \cite{algencan1,algencan2}.  Furthermore, 
this numerical approach is general and may be readily applied to generate other spectral distributions (subject to the constraints mentioned above). 
In the remainder of this paper we shall employ a multi-state Markovian fluctuator representation of $1/\omega$ noise together with a zero frequency component $\eta_{os}$ that describes possible sources of heterogeneous dephasing.  This combination is experimentally relevant to a broad range of physical qubits. 

\begin{figure}[htbp]
	\begin{center}
	\includegraphics[width=\columnwidth]{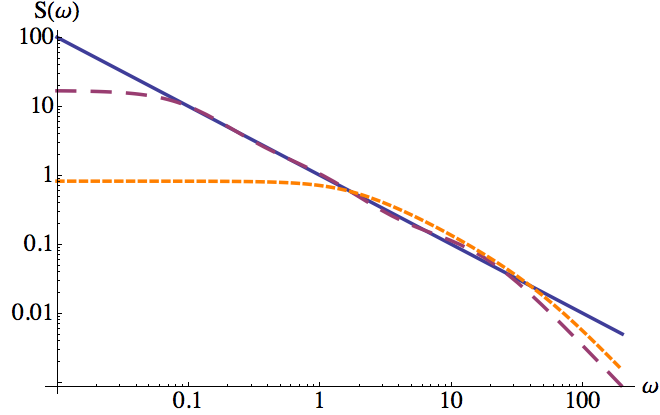}
	\caption{(Color online) Numerically optimized noise power spectral
		density (red, dashed line) with optimization constructed to match the target noise spectrum
		$S_t \propto 1/\omega + \eta_{\rm{os}}^2 \delta(\omega)$  (blue, solid line) over two decades of 
		frequency $\omega$. Also shown 
		is the fit obtained with the analytic representation of Ref.~\cite{kuop2008} using 32 noise states
		(yellow, dotted line).}
	\label{specFit}
	\end{center}
\end{figure}

To illustrate the efficiency and flexibility of this generation of arbitrary noise spectral densities by numerical optimization of a multi-state Markovian fluctuator, we applied the approach to calculation of a 4-state Markovian fluctuator representation of a target spectrum 
$S_t \propto 1/\omega + \eta_{\rm{os}}^2 \delta(\omega)$.  The result is
	\begin{equation}
	\label{etavec}
		\vec\eta = \eta_{\rm{os}} + \epsilon (-0.875,1.36,-1.36,0.875)
	\end{equation}
	\[ \Gamma =  \Gamma_m
		\left(\begin{array}{cccc}
			-7.69   &  7.64   & 0.0322  &   0.0123\\
			7.64    & -8.41    & 0.694 &    0.0694\\
			0.0322 &   0.694  &   -0.730 &    0.00437\\
			0.0123   &    0.0694 &    0.00437 &   -0.0861
		  \end{array}\right). \]
with $\eta_{\rm{os}}$ the constant noise offset responsible for zero frequency noise, $\epsilon$ the scaling of the noise amplitude (represented as a fraction of the maximum control amplitude)
and $\Gamma_m$ a constant that tunes the range of frequencies over which the fluctuator best approximates $S_t$.  
The resulting power spectrum for parameter set $\epsilon = 10^{-3}$, $\Gamma_m = 1/30$ and $\eta_{\rm{os}} = 0$ is shown for a range of finite frequency $\omega$ in Fig.~\ref{specFit}. For comparison we also show the corresponding approximation to the $1/\omega$ power spectrum derived from a 32-state Markovian fluctuator with the analytic form of Ref.~\cite{kuop2008}.    It is evident that the numerically optimized 4-state Markovian fluctuator provides a significantly improved fit relative to the analytic approximation, as well as a significantly greater range of representation.  Such enhanced accuracy, together with the considerable increase in efficiency and greater flexibility, illustrated here by the addition of the zero frequency noise component in the fit (see Eq.~\eqref{etavec}), render this numerical optimization approach to generation of arbitrary spectral noise densities extremely attractive.
  
\section*{Qubit Evolution with Noise}
The evolution of this one qubit system under classical dephasing noise $\eta(t)$ is exactly solvable through the use of conditional density matrices, $\rho_k(t)$, as described in \cite{saira2007}.  We outline here a slightly modified version of this approach.  Defining $\rho_k(t)$ as the density operator of the system conditioned on the environment being in the state $k$,  the total density operator of the system is given by the sum of the conditional density operators weighted by the probability of occupation of the associated noise state:
	\[ \rho(t) = \sum_{j=1}^N p_k(t) \rho_k(t). \]
We choose as the initial probability vector $\vec{p}(0)$ the stationary probability, $\vec p_s$.  For a single qubit, we can
parameterize the conditional density matries $\rho_k(t)$ by their Bloch vectors, $\vec\zeta_k(t)$, 
	\[ \rho_k(t) = \frac{1}{2}\left(\mathcal{I} + \vec\zeta_k(t) \cdot \vec\sigma \right), \]
where $\mathcal{I}$ is the identity operator and $\vec\sigma$ is the vector of Pauli spin-1/2 matrices.  
The resulting stochastic Liouville equation for the conditional density matrices can be transformed to yield the dynamics of the conditional Bloch vectors, which are given by
	\begin{equation}
	\label{zetaEvolve}
		 \frac{d}{dt} \vec\zeta_k(t) = M_k(t) \vec\zeta_k(t) + \sum_j \Gamma_{kj} \vec\zeta_j(t). 
	\end{equation}
Here $M_k \in \mathfrak{so}(3)$ is the generator of  Bloch vector rotations,
	\[ M_k(t) = \left(\begin{array}{ccc} 
		0 & -\eta_k & a_y(t) \\
		\eta_k & 0 & -a_z(t) \\
		-a_y(t) & a_z(t) & 0 
	 \end{array}\right). \]
The second term in Eq.~\eqref{zetaEvolve} describes the effect of the noise switching on the conditional Bloch vectors. We have thus arrived at a set of $N$ coupled matrix differential equations for the evolution of the $N$ conditional Bloch vectors.  These can be solved by treating the set of Bloch vectors as a single vector composed by stacking the conditional Bloch vectors to get a single $3N$-dimensional vector,
	$ \vec Z(t) = \bigoplus_{k=1}^N \vec\zeta_k(t).$
The equation of motion for $\vec Z(t)$ can be straightforwardly derived from Eq.~\eqref{zetaEvolve} and is given by
	\[ \dd{t} \vec Z(t) = \left( \bigoplus_{k=1}^N M_k(t) + \Gamma \otimes \mathcal{I} \right)\vec{Z}(t) \equiv \mathcal{L}(t)\vec Z(t). \]
This is solved formally in the usual way, namely as $\vec Z(t) = \timeorder\exp\left(\int_0^t\mathcal{L}(t^\prime)dt^\prime \right) \vec Z(0)$, where the symbol, $\timeorder$, is the usual Dyson time-ordering operator.  This time-ordered integral becomes a time-ordered product if we restrict the control functions, $a_x(t)$ and $a_y(t),$ to those that are piecewise-constant in time.  The Lindblad operators, $\mathcal{L}(t)$, are also then piecewise constant, taking values $\mathcal{L}_i$ for times $\delta_i$. 
For future convenience, we divide the control functions into $2N_{\rm{p}}$ subintervals, where subinterval $i$ will in general take nonzero amplitude for $i$ even (and be called a "pulse"), and will take zero amplitude for $i$ odd, corresponding to a quiescent time between pulses. Thus, $N_{\rm{p}}$ is understood to mean the number of pulses in the control pulse sequence. 
Each control has time duration $\delta_i$ and the total time for a pulse sequence is equal to $\tau  =\sum_i \delta_i$.
With $2N_{\rm{p}}$ control function values, the corresponding Bloch vector dynamics are given by
	\[ \vec Z(\tau) = \left(\timeorder\prod_{i=1}^{2N_{\rm{p}}} \exp\left(\mathcal L_i \,\delta_i\right) \right) \vec Z(0).\]
	  Calculating the evolution of a given initial state is then a matter of matrix multiplication. Because the probability vector $\vec p(t) = \vec p_s$, all noise states are equally probable and the relation between $\vec Z(t)$ and the Bloch vector, $\vec\zeta(t)$, is given by 
	\[ \vec\zeta(t) = \frac{1}{N}\left(\begin{array}{ccccccc} 1&0&0&1&0&0&\cdots \\0&1&0&0&1&0&\cdots\\0&0&1&0&0&1&\cdots   \end{array}\right)\vec Z(t)
		\equiv \frac{1}{N} \mathcal{I}_N \vec Z(t).\]
The inverse relation is simply $\vec Z(0) = \mathcal{I}_N^\dagger \vec\zeta(0)$ 
and the final Bloch vector is then $\vec\zeta(\tau) = \mathcal{E} \vec \zeta(0)$, where
	\begin{equation}
		\mathcal{E}  \equiv \frac{1}{N} \mathcal{I}_N \cdot  \left(\timeorder\prod_{i=1}^{N_{\rm{p}}} \exp\left(\mathcal L_i \,\delta t_i\right) \right)\cdot  \mathcal{I}_N^\dagger.
	\end{equation}
\section*{Numerically Optimized Control}
The control functions must now be chosen to realize a desired target operation on the Bloch vector, $\vec\zeta \rightarrow  G \vec\zeta$.  We choose the operator fidelity to measure the efficacy of these control functions.  We define the fidelity function as \cite{mikeandike},
	\begin{align*} 
		\Phi_G[a_x(t),a_y(t)] 
			&= \frac{1}{2}\left(1 + \vec \zeta(\tau) \cdot G\, \vec \zeta(0) \right) \\
			&= \frac{1}{2}\left(1 + \mathcal{E}\, \vec \zeta(0) \cdot G\, \vec \zeta(0) \right). 
	\end{align*}
Note that the fidelity is a functional of the control fields, $a_{x/y}(t)$.
From the perspective of quantum information, no state is any more important that any other, so we would ideally like our pulse sequences to maximize the worst-case fidelity over all possible initial states.
However, the minimization over initial states to find the worst-case fidelity is too expensive a computation to yield a useful cost function.  
Therefore we use instead as our cost function the average case fidelity and compare this with the worst case fidelity obtained from the optimized pulse sequence at the end of the computation in order to ascertain the range of errors.  Thus we
	\begin{align*}
		\underset{ a_x(t), a_y(t) }{\mathrm{maximize}} & \hspace{.5cm} \expect{ \,\Phi_G[a_x(t), a_y(t)]\, }_{\vec\zeta(0)}\\
		\mathrm{subject \,to} & \hspace{.5cm} a_x(t)^2 + a_y(t)^2 \le 1,
	\end{align*}
where the notation, $\expect{\cdot}_{\vec\zeta}$, implies an average taken over the surface of the Bloch sphere.  The above average can be evaluated as
	\begin{align*} 
		\expect{ \,\Phi_G[a_x(t), a_y(t)]\, }_{\vec\zeta(0)} 
			&= \frac{1}{2} + \frac{1}{2}\sum_{i,j,k} \expect{ \mathcal{E}_{ij} G_{ik} \zeta_j(0) \zeta_k(0)}_{\vec\zeta{(0)}} \\
			&= \frac{1}{2} + \frac{1}{2}\sum_{i,k}  \mathcal{E}_{ik} G_{ik} \expect{ \zeta_k(0)^2}_{\vec\zeta{(0)}} \\
			&= \frac{1}{2} + \frac{1}{6}\, \tr\!\! \left(\mathcal{E} G^{\mathsf{T}} \right).
	\end{align*}
We have used the result that the average over the surface of a unit sphere is given by $\expect{\zeta_i \zeta_j}_{\vec\zeta} = \delta_{ij}/3$.
 Finally, as discussed above, we demand that the optimized pulse sequences be insensitive to zero-frequency noise. 
This is achieved by including a constant offset value, $\eta_{\rm{os}}$, to the noise vector, as in Eq.~\ref{etavec}.  We systematically analyze the effect of this additional zero frequency noise by choosing the offset to take values within the range $\abs{\eta_{\rm{os}}} \le 10\epsilon$.
  The optimization problem then becomes
	\begin{align}
		\underset{ a_x(t), a_y(t) }{\mathrm{maximize}} & \hspace{.5cm} \underset{\eta_{\rm{os}}}{\rm{min}}\hspace{.2cm} \tr\!\! \left(\mathcal{E} G^{\mathsf{T}} \right)\label{optimization}\\
		\mathrm{subject \,to} & \hspace{.5cm} a_x(t)^2 + a_y(t)^2 \le 1.\notag
	\end{align}

It is in general possible to find analytic gradients of  $\Phi_G[a_x(t),a_y(t)]$ in terms of the pulse sequence parameters by straightforward methods of \cite{najfeld1995} when $\eta_{os}$ is fixed. However, because the objective function has the form of a minimum value over some range of $\eta_{os}$, the objective function is not in general differentiable everywhere. We therefore use the solver \cite{algencan1,algencan2} which employs finite difference approximations to the gradient (which may be undefined in certain regions). A finite difference minimization approach requires many more function evaluations than an explicit gradient calculation, greatly increasing optimization time.

The optimization is performed by undertaking a sampling over the allowed parameter space. We begin by randomly selecting an initial point in the space, and applying numerical optimization techniques to find a locally optimal value of the objective function. We repeat this process many times, each time obtaining a value for $\Phi_G$. After some fixed number of initial conditions are sampled (typically thousands), the pulse sequence obtaining the greatest value of the operator fidelity, $\Phi_G$, is selected as the optimal sequence.

It is important to note that the dimension of the parameter space for optimization is $4N_{\rm{p}}$. A pulse sequence contains $N_{\rm{p}}$ pulses, each of which can be characterized by four parameters: the amplitude, duration, phase angle of the control fields, and quiescent period before the next pulse. To thoroughly sample the enlarging parameter space, the number of initial points sampled for optimization should  grow as $n_s^{4N_{\rm{p}}}$, where $n_s$ is the number of initial samples taken for $N_{\rm{p}} = 1$.

\begin{figure}[t] 
 \includegraphics[width=3.4in]{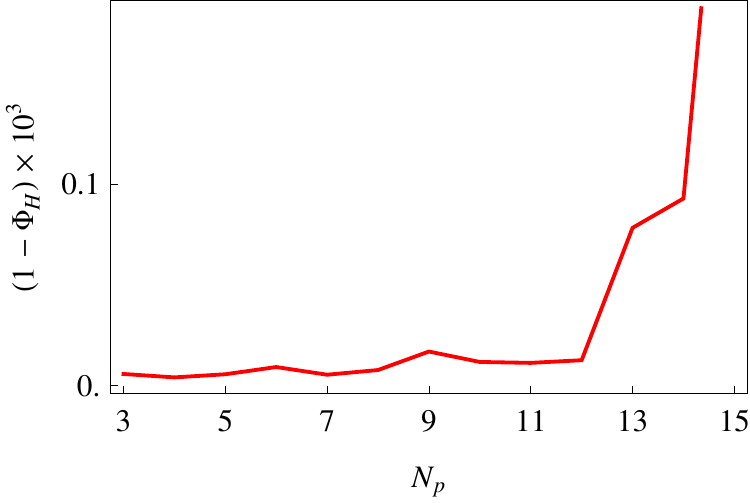}
\caption{ Worst-case error, $(1-\Phi_{{H}})$, over the Bloch sphere for a Hadamard gate generated by GRAPE as a function of the number of pulses in the pulse sequence up to $N_p = 15$ for fixed computational effort.  See text for disucssion.}
\label{fig:hnp}
\end{figure}

An interesting point is raised by Fig.~\ref{fig:hnp}, which shows that the worst-case error of a numerically optimized Hadamard gate as a function of $N_{\rm{p}}$.  If the computational effort is allowed to grow exponentially with $N_{\rm{p}}$ (and therefore linearly with the size of the parameter space), we can expect the gate error to decrease monotonically with $N_{\rm{p}}$.  With exponential computational resources, one is able to sample all of the parameter space defining the pulse sequences.   And because the set of $N_{\rm{p}} = n $ pulse sequences is a strict subset of the set of $N_{\rm{p}} = n + 1$ pulse sequences, such a search should yield sequences that, at the very worst, do not decrease in efficacy.   However, when  the number of initial sample points does not grow exponentially with $N_{\rm{p}}$, the performance of GRAPE will suffer because it becomes exceedingly unlikely to find good optima of the objective function as $N_{\rm{p}}$ becomes large. This exponential scaling in the number of sample points required places an upper limit on the largest values of $N_{\rm{p}}$ for which the GRAPE approach will be useful.
In the examples studied in this work, we find that for gates other than the identity, this scaling restricts values of $N_{\rm{p}}$ to a maximum of 6-10.  Thus in Fig.~\ref{fig:hnp} we see that the error begins to rise after $N_{\rm{p}} = 10$.  For non-trivial gates this is not a disadvantage since it is in any case advantageous to implement the gate in as short a time as possible to avoid decoherence.  The exception to this is the identity gate, which one wishes to realize over as long a time as possible when implementing a quantum memory.  In this case we find that good identity gates may be found for $N_{\rm{p}} \sim 30$.

\section{Results}
To demonstrate the power and flexibility of the improved multi-state Markovian fluctuator approach to coherent control of qubit dephasing, we apply it to studying magnetic interface noise experienced by a phosphorus dopant atom implanted in the channel of a silicon MOSFET.   
de Sousa has proposed a model of this noise as caused by  dangling bonds located at the interface between the crystalline Si and the amorphous oxide  \cite{sousa2007}.  These defects, known as $P_b$ centers, are  associated with oxygen vacancies in the oxide and have the structure $\cdot {\rm{Si}}\equiv {\rm{Si}}_3$.  The lone electron in the dangling bond  can provide a thermally switching paramagnetic contribution to the magnetic environment experienced by the donor that causes magnetic field noise.  This noise then acts to dephase the electron spin qubit defined on the phosphorus dopant.  
By modeling the $P_b$ center spin flips as coupled to tunneling two-level systems in the oxide, de Sousa has 
shown that the resulting magnetic field noise possesses an approximate power spectral density $S(\omega) \propto 1/\omega$ \cite{sousa2007}.  Recent work by Paik et al.~has provided evidence in favor of de Sousa's model \cite{paik2010}.

We can empirically estimate the proportionality constant, or equivalently the noise strength $\epsilon$ (Eq.~(\ref{etavec})) by comparing the calculated $T_2$ time of an electron spin qubit with the experimental values extracted for phosphorus donors implanted in Si MOSFET devices. Donor electron $T_2$ times have been shown to be  several milliseconds in isotopically purified silicon \cite{schenkel2006, tyryshkin2003}.

We can determine an effective $T_2$ time for a pulse sequence in our model as follows. We initialize a qubit in the $+1$ eigenstate of $\sigma_x$, and apply a quantum memory pulse sequence repeated several times. If one measures only at the conclusion of each repetition of the pulse sequence, the quantity $\langle \sigma_x \rangle$ will decay approximately exponentially in time, with a time constant that we define to be $T_2$.

\begin{figure}[t] 
 \includegraphics[width=\columnwidth]{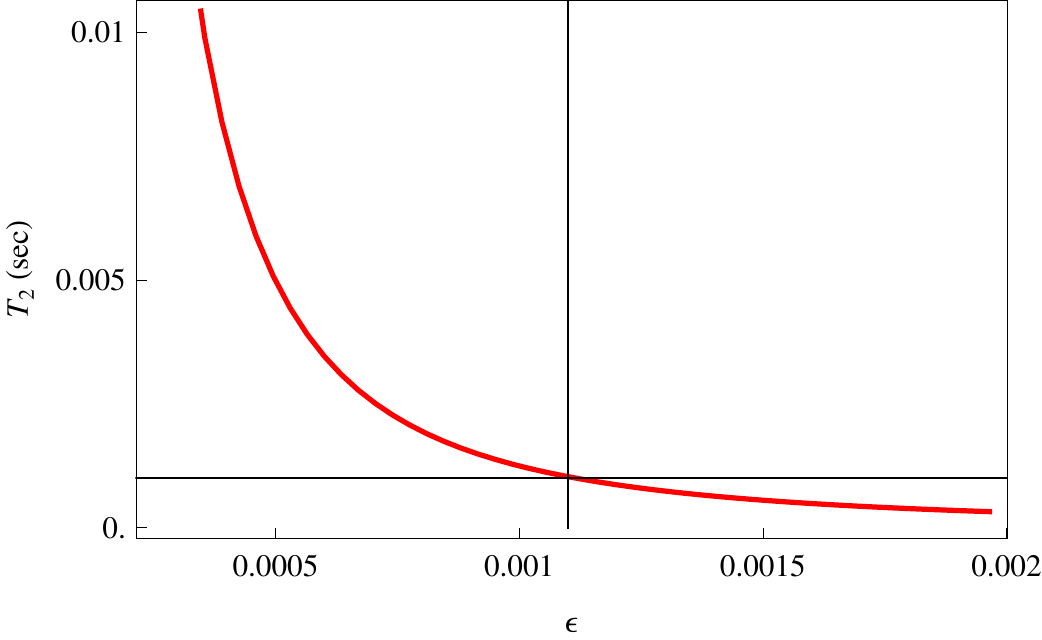} 
\caption{(Color online) Calculated qubit $T_2$ times as a function of the noise parameter $\epsilon$. Results are presented for qubit $T_2$ with the application of a Carr-Purcell sequence with a 1\% duty cycle. The intersecting horizontal and vertical lines indicate the value of $\epsilon$ at which the electron $T_2$ equals $1 \rm{ms}$.
}
\label{fig:t2}
\end{figure}

We have set the value of the parameter $\epsilon$ to be that which yields $T_2 \approx 1 \mathrm{ms}$ under a Carr-Purcell pulse sequence operated with a $1\%$ duty cycle. As demonstrated in Figure \ref{fig:t2}, this condition is satisfied with $\epsilon = 0.0011$, which we have rounded to $ \epsilon = 10^{-3}$.

In the remainder of this section we explore the power of numerically optimized pulse sequences obtained with the improved multistate Markovian fluctuator, for two target unitary operations subject to this interface-induced $1/\omega$ noise with an additional constant noise offset that allows for heterogeneous dephasing.   The first is quantum memory, i.e., the preservation of coherence of an arbitrary quantum state, while the second is a single-qubit Hadamard transformation.  We find that excellent performance of GRAPE for both operations can be attained even in the presence of additional zero-frequency (constant) noise.

Consistent with the application to experimentally accessible phosphorus dopants silicon devices, we construct here pulse sequences that may be implemented by current signal generators and that are thus subject to limitations on the on/off ratio. Consequently, we  enforce a 50\% duty cycle on the GRAPE sequences, i.e., each pulse is followed by a brief quiescent period and where the total quiescent time must be at least half of the total pulse length, or, $\tau/2$.

\subsection{Quantum Memory}
We begin this section with a discussion of coherence maintaining operations, known generally as \emph{quantum memory gates}.  When designing such pulses, one must make pulse design decisions based on the specifics of the experiment in question.  Consider a particular experiment which requires that coherence be maintained for a certain time, $t$.  Ideally, one would design a pulse sequence itself having total length $t$, as well.  In principle, such a pulse sequence would have much more flexibility than a sequence of duration $t/n$ repeated $n$ times.  However, the computational effort scales exponentially in the number of pulses, as discussed above, and long sequences may be difficult to find which match or exceed the performance of repeated short sequences.  Which choice is made will depend strongly on the computational resources available to the pulse designer.  
 
\begin{figure}[htbp]
	\begin{center}
	\includegraphics[width=3.5in]{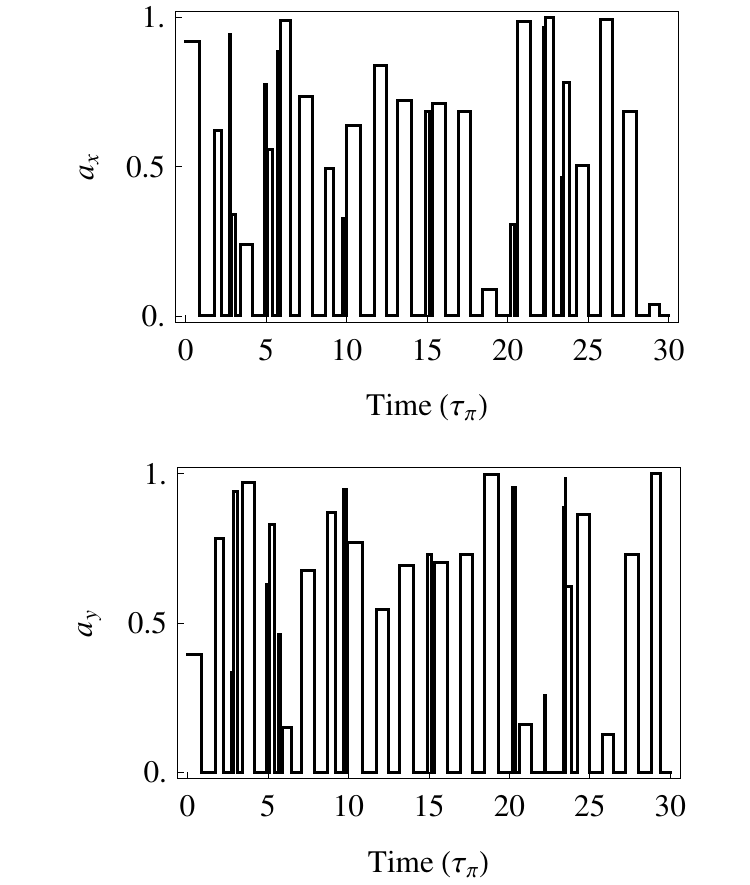}
	\caption{(a) $x$-axis control function and (b) $y$-axis control function for implementing quantum memory operations.}
	\label{fig:identity}
	\end{center}
\end{figure}

Here we present a numerical solution of Eq.~\eqref{optimization} with $G$ the identity matrix, total pulse sequence time $\tau = 30 \tau_\pi$ and total number of pulses $N_{\rm p} = 30$.  The particular optimal solution found under these constraints is shown in Fig.~\ref{fig:identity}.
Here $\tau_\pi$ refers to the time required to perform a full $\pi$ rotation of the qubit at maximum control amplitude. This value of $N_{\rm p}$ was chosen because it was the largest value for which we were able to obtain results in a reasonable amount of computer time (see discussions above).  
To compare with these numerically optimized pulse sequences we  construct an equivalent length Carr-Purcell (CP) decoupling pulse sequence, defined by
\[
w-\pi_x - w - w - \pi_x - w
\] repeated 7 times, where $\pi_x$ denotes a $\pi$-pulse about the $x$-axis, and $w$ denotes a quiescent period of $\tau_w = \frac{4}{7}\tau_\pi$.  

\begin{figure}[htbp]
\begin{center}
\includegraphics[width=3in]{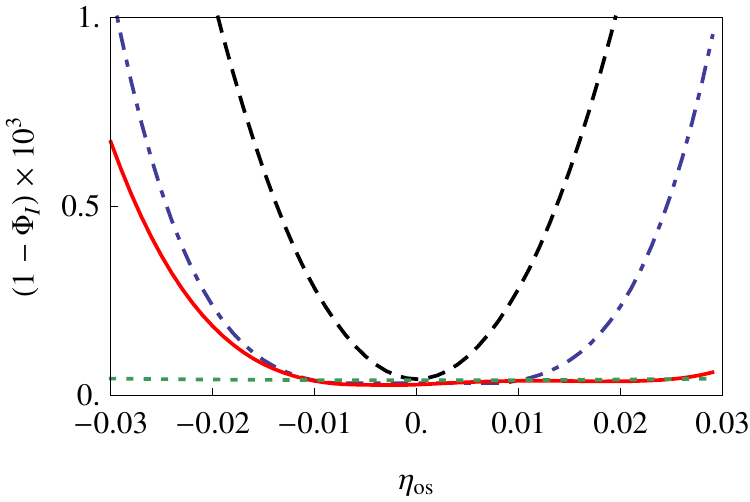}

\caption{ (Color online) 
Error ($1-\Phi_{{I}}$) in quantum memory of a qubit subject to dephasing noise with spectral density $S_t \propto 1/\omega + \eta_{\rm{os}}^2 \delta(\omega)$ under various control pulse sequences of duration $30\tau_{\pi}$, shown as a function of offset noise values $\eta_{os}$. The solid red line represents the error for a sequence that is optimized over a range of offset noise. 
Green dotted line: error obtained with  infinite-amplitude Carr--Purcell sequence. Blue dot-dashed line: error obtained with finite amplitude Carr-Purcell sequence.  Also shown as the black, dashed line is a pulse sequence obtained without regard to robustness over $\eta_{\rm{os}}$.}

\label{eos}
\end{center}
\end{figure}

Fig.~\ref{eos} shows the error (defined as 1 - $\Phi_{{I}}$) as a function of the zero frequency noise $\eta_{\rm{os}}$ for a numerically generated pulse sequence that is optimized over all values of $\eta_{os}$ (red line), in addition to optimization against the $1/\omega$ noise.  The blue and green lines show the corresponding fidelities obtained with the Carr--Purcell sequence of equivalent duration specified above, using finite amplitude 
(dot-dashed blue line) and infinite amplitude (dotted green line) pulses.  

Infinite amplitude Carr--Purcell pulses are capable of refocusing arbitrarily large zero-frequency noise, resulting in a
constant error as a function of $\eta_{\rm{os}}$ whose value can be taken as a measure of the uncorrected error due to the $1/\omega$ noise component.  Unlike the ideal, infinite amplitude pulse sequence, a Carr--Purcell sequence with finite amplitude pulses does not allow complete Bloch sphere rotations, which prevents the  exact refocusing of zero-frequency noise, resulting in a significant rise in error as the zero-frequency noise magnitude $|\eta_{os}|$ increases.  Note that the lack of time reversal symmetry possessed by the optimized pulses (unlike the Carr--Purcell sequence) results in an asymmetry with respect to $\eta_{\rm{os}}$, as illustrated in Fig.~\ref{eos}.  

The numerically optimized sequence shows improved performance relative to these Carr-Purcell decoupling sequences in two respects.  First, it performs better than the Carr--Purcell pulse sequences for zero and small $|\eta_{os}|$, due to the greater flexibility of the numerical optimization in developing protection against the $1/\omega$ noise component. Thus, at $\eta_{\rm{os}}=0$, the error obtained with the numerically optimized pulse sequence is  $2.88\times10^{-5}$, compared to $3.26\times10^{-5}$ with the finite amplitude Carr--Purcell pulse sequence. 
However, a far more dramatic difference is the greater robustness against the magnitude of zero-frequency noise.  The numerically optimized pulse sequence is seen to show very small error over a broad range of $\eta_{os}$, attesting to the power of the numerical approach to mitigate combined decoherence effects deriving from very different noise sources.

\subsection{Hadamard Gate}
\label{unitaries}
Our second target operation is the Hadamard gate, 
	\begin{equation}
		H = \frac{1}{\sqrt{2}}
		\left( \begin{array}{cc}
			1 & 1 \\
			1 & -1 \\
			\end{array} \right),
	\end{equation}
a common single qubit operation in quantum algorithms. The optimization considerations for implementing such a single qubit rotation with numerically optimized pulse sequences are similar to those for generating sequences to protect the identity gate. However, in contrast to the situation for quantum memory, here we are interested in maximizing fidelity and robustness to a constant noise offset, rather than in the maintenance of coherence over a long time.  Thus the optimal pulse sequences for protection of the Hadamard gate are considerably shorter than the sequences derived above for protection of quantum memory. 

\begin{figure}[htbp]
\begin{center}
\includegraphics[width=3in]{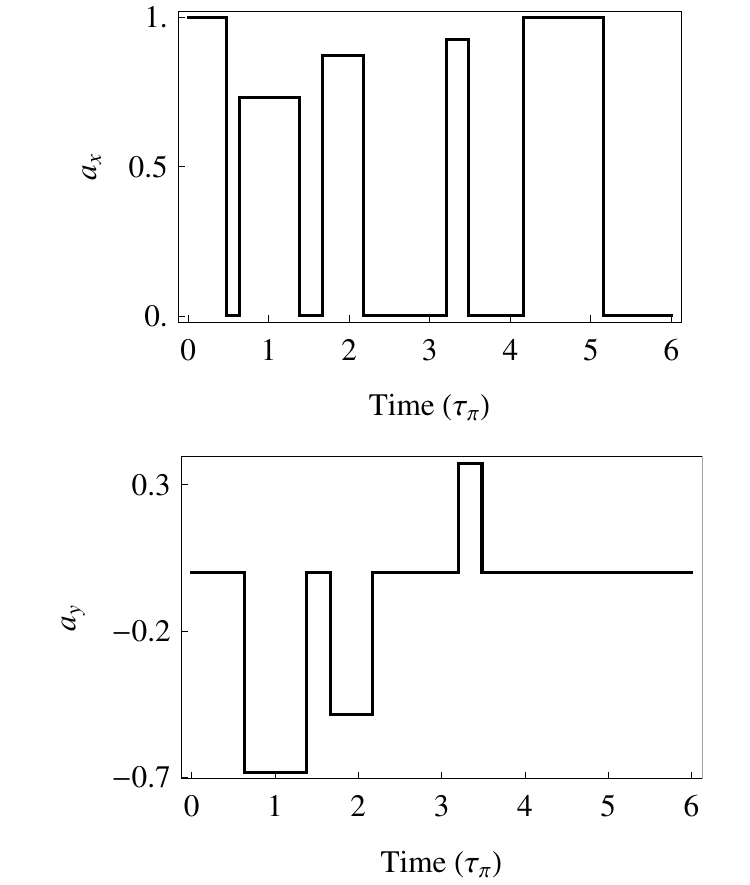}
\caption{Two-dimensional control function producing a high fidelity $H$ rotation in $T=6\pi$ for $\eta_{os}=0$.
 This pulse sequence results in a worst case fidelity of $\Phi_{{H}} = 1 - 8.27\times10^{-6}$ and exhibits a strong robustness to the value of constant offset noise $\eta_{os}$.  Panel (a) shows the $x$-axis control function and panel (b) the $y$-axis control.}
\label{hadamard}
\end{center}
\end{figure}

Using the same cost function as Eq.~\eqref{optimization} and evaluating $\Phi_{{H}}$ for the $H$ operation, we were able to obtain high-fidelity pulse sequences with $\tau = 6\tau_\pi$ and $N_{\rm p} = 6$.  Fig.~\ref{hadamard} shows the resulting pulse sequence when optimization is made for the case of zero offset noise, $\eta_{os}=0$. %

\begin{figure}[htbp] 
\begin{center}
\includegraphics[width=3in]{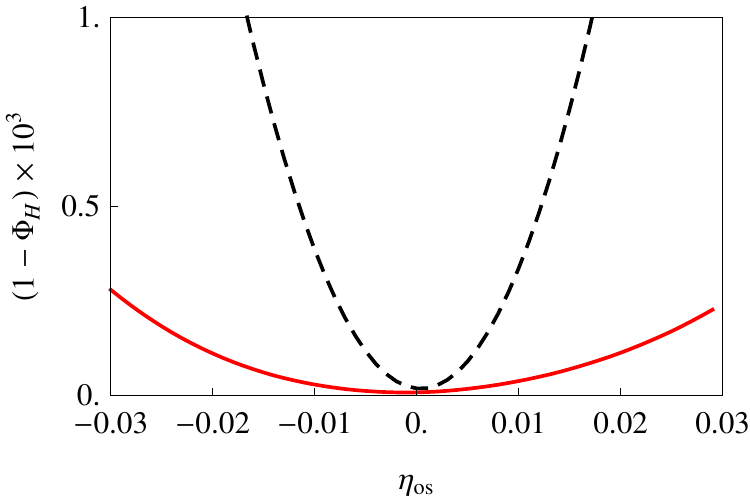}
\caption{(Color online) Error (1-$\Phi_{{H}}$) of a $H$ rotation under  numerically optimized pulse sequences mitigating against noise with spectral density $S_t \propto 1/\omega + \eta_{\rm{os}}^2 \delta(\omega)$ as a function of $\eta_{\rm{os}}$. The solid, red line shows the result of optimizing with respect to $\eta_{\rm{os}}$ as well as the $1/\omega$ noise.  The dashed, black line shows the considerably less robust result of using a pulse sequence that is optimized only at a single value of $\eta$ ($\eta_{\rm{os}} =0$). }
\label{heos}
\end{center}
\end{figure}

This pulse sequence results in a worst case error for the Hadamard gate of $8.27\times10^{-6}$ at 
$\eta_{os}=0$. Thus it is evident that gate operations can readily be corrected at similar or better levels than quantum memory, using shorter pulse sequences.  Fig.~\ref{heos} shows the worst case %
error for a numerically optimized pulse sequence optimized over a range of $\eta_{\rm{os}}$ values, as before, as a function of the noise offset $\eta_{os}$.  Comparison with the results obtained with a single pulse sequence that is optimized only for $\eta_{os} =0$ shows again the enhanced robustness afforded by the numerical optimization approach.

\section{Summary}
We have expanded the techniques of \cite{kuop2008, mott2006, saira2007} to develop a general numerical method for simulating noise sources deriving from a broad variety of Markovian power spectra. The method employs a new numerical approach to generation of the noise spectrum that can greatly reduce the number of noise states required to simulate a noise source with a given spectral density.  We illustrated this with the example of a four state simulation of a $1/\omega$ spectrum over two decades of frequency $\omega$, which is significantly more efficient than the constructive method employed previously in Ref.~\cite{mott2006}.
This numerical representation of Markovian noise was then used in the context of numerical generation of control pulse sequences to mitigate the effects of this noise on a single qubit.  Here we have extended the methods of \cite{mott2006} to allow control pulses to be performed along both $x$- and $y$-axes of the qubit, rather than along a single axis.   
Using numerical generation of the noise allows spectral densities from different sources of dephasing to be combined, giving rise to considerable additional flexibility and robustness in the decoherence mitigation.  This was illustrated by generation of pulse sequences designed to minimize decoherence in the presence of both homogeneous dephasing characterized by $1/\omega$ spectral density and a source of heterogeneous dephasing, characterized by a zero frequency noise offset $\eta_{os}$.  The numerical optimization approach allows the pulse sequences to be simultaneously optimized with respect to the parameter $\eta_{\rm{os}}$ and the $1/\omega$ noise. This introduces an unprecedented robustness to decoherence mitigation with realistic bounded amplitude controls in the presence of resonance frequency errors and inhomogeneous broadening.  In particular, the performance of the numerically optimized pulse sequences over a range of $\eta_{\rm{os}}$  values was seen to be considerably superior to the corresponding performance of
 a standard dynamical decoupling pulse sequence with bounded amplitudes.  
 
As a demonstration of the power and flexibility of these numerical methods for noise mitigation, we have explicitly studied the protection of quantum memory and the protection of the Hadamard gate.  To ground the derived pulse sequences to a physical system, we took estimates of noise strength that are appropriate to the situation of dephasing noise acting on phosphorus donors in silicon and implemented the numerical optimization subject to realistic constraints of duty cycle and pulse amplitude limitations. The remarkable robustness of the optimal pulse sequences with respect to the constant noise offset, showing worst case gate errors of order $10^{-6}-10^{-5}$ over a range of noise offsets, is encouraging for application of these pulse sequences to current experiments with spin qubits in semiconductors \cite{schenkel2006}. 

\section{Acknowledgements} 
This work has been supported by the National Security Agency under MOD713100A.  DJG also thanks UC LEADS for financial support.  The authors would like to thank Thomas Schenkel for many useful discussions.  

\bibliography{siqc_pulse}
\bibliographystyle{apsrev}

\end{document}